\documentclass[preprint]{revtex4-1}
\usepackage[utf8]{inputenc}

\usepackage{graphicx}
\usepackage{amsmath}
\usepackage{comment}
\usepackage[a4paper,top=3cm,bottom=2cm,left=3cm,right=3cm,marginparwidth=1.75cm]{geometry}
\begin{document}
\title{Gravitational waves from binary black holes as probes of \\the structure formation history}
\author{Tomohiro Nakama}
\affiliation{Jockey Club Institute for Advanced Study, The Hong Kong University of Science and Technology, Hong Kong, P.R. China}

\begin{abstract}
Gravitational-wave detectors on earth have detected gravitational waves from merging compact objects in the local Universe.
    In future we will detect gravitational waves from higher-redshift sources, which trace the high-redshift structure formation history. That is, by observing high-redshift gravitational-wave events we will be able to probe structure formation history. This will provide additional insight into the early Universe when primordial fluctuations are generated and also into the nature of dark matter.
\end{abstract}
\maketitle

\section{Introduction}
The first detection of gravitational waves (GWs) from merging black holes (BHs) was reported in \cite{Abbott:2016blz}. The source was inferred to lie at a luminosity distance of $410^{+160}_{-180}$Mpc, and assuming the standard cosmology \cite{Ade:2015xua}, the redshift corresponding to this distance was reported to be $0.09^{+0.03}_{-0.04}$. Since then, the detections of somewhat higher-redshift events have already been reported \cite{ligo}. In future, GWs from even higher-redshift ($z>{\cal O}(1)$) binary BHs are expected to be detected with more sensitive detectors operating at different frequency bands (see e.g. \cite{Evans:2016mbw}). Their redshift distribution was pointed out to help us identify the origin of binary BHs in \cite{Nakamura:2016hna,Koushiappas:2017kqm}. 

A number of physical processes are involved before merging BHs arise in the Universe. First, primordial fluctuations generated in the early Universe later cause fluctuations in the density of matter, and matter density perturbations gradually grow due to gravity. Eventually, matter overdensities collapse to form gravitationally bound objects called halos. In some of these halos, baryons are concentrated so that stars are formed, and some stars leave BHs at the end of their life. Some of these BHs find companions to form gravitationally-bound binaries, and BH binaries shrink due to e.g. GW emission. Eventually, those BHs merge, and some of such mergers take place within the age of the Universe. Some of such mergers have been and will be detected by GW detectors operating on or near earth. These considerations imply that the event rate of BH mergers at high redshifts traces structure formation history at corresponding redshifts. Since structure formation at high redshifts is sensitive to the properties of dark matter (DM) or small-scale primordial fluctuations, high-redshift GW events would allow us to derive constraints on DM physics or small-scale primordial fluctuations. For instance, structure formation is delayed relative to that in cold DM (CDM) scenarios in DM scenarios such as warm DM (see \cite{Bode:2000gq} and references therein) or wave DM ($\psi$DM) \cite{Peebles:2000yy,Hu:2000ke}, or when small-scale primordial fluctuations are suppressed on small scales \cite{Kamionkowski:1999vp}. As discussed in \cite{Nakama:2017ohe}, such scenarios have been studied in the context of the so-called $\Lambda$CDM small-scale crisis, i.e. discrepancies between predictions and observations about small-scale structures of the Universe. We will illustrate how we can probe high-redshift structure formation from GW events by taking the CDM scenario and a $\psi$DM scenario as examples, and our work will for instance have implications for the small-scale crisis. 

\section{Quantifying Star Formation History}
Estimating the rate of BH mergers involves quantifying star formation history, which depends on halo formation history. We use HMFcalc \cite{Murray:2013qza} for the halo mass function $dn(M,z)/d\mathrm{ln}M$, which is the comoving number density of halos per unit logarithm of halo mass $M$. 

As mentioned in the Introduction, we use a wave or fuzzy DM (see \cite{Niemeyer:2019aqm} for overview) as an example which predicts structure formation histories  largely different from those in CDM scenarios. In a $\psi$DM scenario, DM is assumed to be extremely light bosons, such as axion-like particles predicted in string theory \cite{Arvanitaki:2009fg}. One important feature of such a scenario is suppression of structure formation on small scales due to the uncertainty principle, where the characteristic scales below which structure formation is suppressed are determined by the nature of $\psi$DM such as its mass (see \cite{Niemeyer:2019aqm} and references therein). Possibilities of $\psi$DM have recently attracted increasing attention partly because more and more tight experimental limits have been obtained on popular DM candidates such as weakly-interacting-massive particles \cite{Arcadi:2017kky}, and also because $\psi$DM has the potential to help alleviate or solve the  $\Lambda$CDM small-scale crisis \cite{DelPopolo:2016emo}. 

For $\psi$DM, the halo mass function is suppressed on small scales as indicated by \cite{Schive:2015kza} 
\begin{equation}
    \frac{dn(M,z)}{dM}\bigg|_{\psi \mathrm{DM}}=\frac{dn(M,z)}{dM}\bigg|_{\mathrm{CDM}}\left[1+\left(\frac{M}{M_0}\right)^{-1.1}\right]^{-2.2},
\end{equation}
where $M_0=1.6\times 10^{10}m_{22}^{-4/3}M_\odot$ and  $m_{22}=m_\psi/(10^{-22}\mathrm{eV})$ ($m_\psi$ is the mass of $\psi$DM). The above fitting function describes the mass functions obtained by simulations of structure formation of $\psi$DM universes well (see Fig. 4 of \cite{Schive:2015kza}). 

The fraction of baryons in structures which are more massive than $M_{\mathrm{min}}$ is \cite{Tan:2016xvl}
\begin{equation}
    f_b(z)=\rho_{m,0}^{-1}\int_{M_{\mathrm{min}}}^\infty\frac{dn(M,z)}{d\mathrm{ln}M}dM,
\end{equation}
where $\rho_m$ denotes the matter energy density and the subscript \textquotedblleft0\textquotedblright \, here and hereafter implies quantities at present. 
We take into account halos whose virial temperature is larger than $10^4$ K, i.e., $M_{\mathrm{min}}=10^9(1+z)^{-3/2}h^{-1}M_\odot$ \cite{Schneider:2001bu}. 
The comoving energy density of baryons accreting onto structures per unit cosmic time is \cite{Daigne:2005dp,Tan:2016xvl}
\begin{equation}
    a_b(t)=\rho_{b,0}\left(\frac{dt}{dz}\right)^{-1}\frac{df_b(z)}{dz},
\end{equation}
where $\rho_{b,0}$ is the current baryon energy density. 
The age of Universe and redshift are related via 
\begin{equation}
    \frac{dt}{dz}=\frac{-H_0^{-1}}{(1+z)\sqrt{\Omega_\Lambda+\Omega_m(1+z)^3}}.
\end{equation}
We use the following expression to estimate the star formation rate (SFR), the total mass of stars arising per unit time per comoving volume, for Pop I/II stars \cite{Daigne:2005dp}:
\begin{equation}
    \Psi_{\mathrm{I}\hspace{-.1em}/\hspace{-.1em}\mathrm{I}\hspace{-.1em}\mathrm{I}}(t)=f_1\frac{\rho_g(t)}{t_1}e^{-(t-t_{i})/t_1}(1-e^{-Z_{\mathrm{IGM}}/Z_{\mathrm{crit}}}),
\end{equation}
where $\rho_g$ is the comoving gas energy density in structures and we set $t_{1}=3.8 $ Gyr and $f_1=0.83$ \cite{Tan:2016xvl}. $t_i$ is the moment when stars start to form, and we assume  $z_i=30$ for the corresponding initial redshift \cite{Tan:2016xvl}. $\rho_g$ is determined by solving an evolution equation presented later (Eq. (\ref{rhog})), which takes into account accretion of baryons onto structures, star formation, stellar evolution and baryon outflow from structures to the intergalactic medium (IGM). $Z_{\mathrm{IGM}}$ is the metallicity of the IGM to be calculated by Eq. (\ref{zigm}), also shown later. We set $Z_{\mathrm{crit}}=10^{-3.5}Z_\odot$ \cite{Tan:2016xvl} with $Z_\odot=0.02$ \cite{Kampakoglou:2007np}. For Pop III stars we use \cite{Daigne:2004ga,Daigne:2005dp,Tan:2016xvl}
\begin{equation}
    \Psi_{\mathrm{I\hspace{-.1em}I\hspace{-.1em}I}}=f_2\frac{\rho_g(t)}{t_2}
    e^{-Z_{\mathrm{IGM}}/Z_{\mathrm{crit}}},
\end{equation}
where $f_2=0.045$ and $t_2=50$ Myr \cite{Daigne:2004ga}. The gas mass converted into stars per comoving volume per unit time is 
\begin{equation}
    \frac{d^2M_\star}{dVdt}=\Psi_{\mathrm{I\hspace{-.1em}/\hspace{-.1em}I\hspace{-.1em}I}}(t)+\Psi_{\mathrm{I\hspace{-.1em}I\hspace{-.1em}I}}(t).
\end{equation}
We assume the Salpeter initial mass function (IMF) \cite{Salpeter:1955it} for both Pop I/II and Pop III stars, which is $\Phi(m)\propto m^{-(1+x)}, (x=1.35)$. The normalization is \cite{Daigne:2005dp}
\begin{equation}
    \int_{m_{\mathrm{inf}}}^{m_{\mathrm{sup}}}m\Phi(m)dm=1,
\end{equation}
where $m_{\mathrm{inf}}=0.1M_\odot(100M_\odot)$ and $m_{\mathrm{sup}}=100M_\odot(500M_\odot)$ for Pop I/II (III) stars \cite{Tan:2016xvl}. 

Part of masses in stars with $M_\odot<m<260 M_\odot$ \cite{Tan:2016xvl} gets converted back to gas mass due to stellar evolution. To quantify this effect, let us define
\begin{equation}
    \Pi_1(m,t) = \begin{cases}
    [m-m_r(m)]\Phi_{\mathrm{I\hspace{-.1em}/\hspace{-.1em}I\hspace{-.1em}I}}(m)\Psi_{\mathrm{I\hspace{-.1em}/\hspace{-.1em}I\hspace{-.1em}I}}[t-\tau(m)] & (1M_\odot<m<100M_\odot) \\
    [m-m_r(m)]\Phi_{\mathrm{I\hspace{-.1em}I\hspace{-.1em}I}}(m)\Psi_{\mathrm{I\hspace{-.1em}I\hspace{-.1em}I}}[t-\tau(m)] & (100M_\odot<m<260M_\odot)
  \end{cases}
\end{equation}
Here, $m_r(m)$ is the remnant mass after stellar evolution, for which we use relations found in \cite{Tan:2016xvl}. $\tau(m)$ is the stellar lifetime, and we use \cite{Madau:2000pa}
\begin{equation}
    \frac{\tau(m)}{\mathrm{Myr}}=\begin{cases}
    33\left(\frac{m}{8M_\odot}\right)^{-3/2} & (m<28.4M_\odot) \\
    3.4\left(\frac{m}{60M_\odot}\right)^{-1/2} & (m>28.4M_\odot)
  \end{cases}
\end{equation}
The extrapolation of this relation to high masses was noted there to be uncertain, but the above formula predicts lifetimes for Pop III stars comparable to what are shown in \cite{Schaerer:2001jc}. Then, the amount of mass converted from stars to gas per unit time per unit comoving volume is
\begin{equation}
    \frac{d^2M_{\mathrm{ej}}}{dVdt}=
    \begin{cases}
    0 & (t_i<t<t_i+\tau_{260}) \\
    \int_{m_i(t)}^{260M_\odot}\Pi_1(m,t)dm & (t_i+\tau_{260}<t<t_i+\tau_{1}) \\
    \int_{M_\odot}^{260M_\odot}\Pi_1(m,t)dm & (t_i+\tau_{1}<t)
  \end{cases}
\end{equation}
where $\tau_{260}=\tau(260M_\odot)$, $\tau_{1}=\tau(M_\odot)$ and $\tau[m_i(t)]=t-t_i$. 

Next, let us quantify the outflow of baryons from structures to IGM. We introduce
\begin{equation}
    \Pi_2(m,t) = \begin{cases}
    2\epsilon v_{\mathrm{esc}}^{-2}(z)\Phi_{\mathrm{I\hspace{-.1em}/\hspace{-.1em}I\hspace{-.1em}I}}(m)\Psi_{\mathrm{I\hspace{-.1em}/\hspace{-.1em}I\hspace{-.1em}I}}[t-\tau(m)]E_{\mathrm{I\hspace{-.1em}/\hspace{-.1em}I\hspace{-.1em}I}}(m) & (8M_\odot<m<100M_\odot) \\
    2\epsilon v_{\mathrm{esc}}^{-2}(z)\Phi_{\mathrm{I\hspace{-.1em}I\hspace{-.1em}I}}(m)\Psi_{\mathrm{I\hspace{-.1em}I\hspace{-.1em}I}}[t-\tau(m)]E_{\mathrm{I\hspace{-.1em}I\hspace{-.1em}I}}(m) & (100M_\odot<m<260M_\odot)
  \end{cases}
\end{equation}
where we use $\epsilon=10^{-3}$ \cite{Tan:2016xvl}, 
\begin{equation}
    v_{\mathrm{esc}}^2(z)=\frac{\int_{M_{\mathrm{min}}}^\infty dM[dn(M,z)/d\mathrm{ln}M][2GM/R(M,z)]}{\int_{M_{\mathrm{min}}}^\infty dM[dn(M,z)/d\mathrm{ln}M]},
\end{equation}
\begin{equation}
    R(M,z)=\left(\frac{3M}{178\rho_{c,0}[\Omega_{m}(1+z)^3+\Omega_{\Lambda}]4\pi}\right)^{1/3},
\end{equation}
and $E_{\mathrm{I\hspace{-.1em}/\hspace{-.1em}I\hspace{-.1em}I}}(m)=10^{51}$ erg and $E_{\mathrm{I\hspace{-.1em}I\hspace{-.1em}I}}(m)=10^{52}$ erg. See \cite{Tan:2016xvl,Kampakoglou:2007np} for the details. Then the amount of mass ejected from structures to the IGM per comoving volume per unit time is
\begin{equation}
    o(t)=
    \begin{cases}
    0 & (t_i<t<t_i+\tau_{260}) \\
    \int_{m_i(t)}^{260M_\odot}\Pi_2(m,t)dm & (t_i+\tau_{260}<t<t_i+\tau_{8}) \\
    \int_{8M_\odot}^{260M_\odot}\Pi_2(m,t)dm & (t_i+\tau_{8}<t)
  \end{cases}
\end{equation}
where $\tau_8=\tau(8M_\odot)$. We estimate $Z_{\mathrm{IGM}}$ by \cite{Tan:2016xvl}
\begin{equation}
    Z_{\mathrm{IGM}}=\frac{\int_{t_i}^t o(t)dt}{\rho_{b,0}+\int_{t_i}^t[o(t)-a_b(t)]dt}. \label{zigm}
\end{equation}

Finally, $\rho_g$ is determined by solving the following equation \cite{Tan:2016xvl}:
\begin{equation}
    \dot{\rho}_g=-\frac{d^2M_\star}{dVdt}+\frac{d^2M_{\mathrm{ej}}}{dVdt}+a_b(t)-o(t). \label{rhog}
\end{equation}
For the initial value, we assume $\rho_g(t_i)=\rho_{b,0}f_b(z_i)$. SFR as a function of $z$ is shown in Fig. 1. For $\psi$DM star formation is significantly suppressed at high redshift. This is expected to suppress the BH-BH merger rate at high redshift as shown later. 

Here we would like to emphasize that, though calculations of SFR as a function of redshift have already been discussed by different authors, they are in most cases done assuming CDM, and nearly-scale-invariant Gaussian primordial fluctuations, except for relatively a small number of works. Since structure formation depends on the nature of DM as well as the nature of primordial fluctuations (their power spectrum and possible deviations from Gaussianity), SFR depends on them, too, as illustrated by our calculations based on a $\psi$DM scenario as an example. That is, SFR encodes information about the nature of DM and primordial fluctuations, and hence observations linked to SFR encode information about those too. One of such observations is the redshift dependence of GW events, which we illustrate in the next section. 
\begin{figure}[htbp]
\begin{center}
  \includegraphics[clip,width=10.0cm]{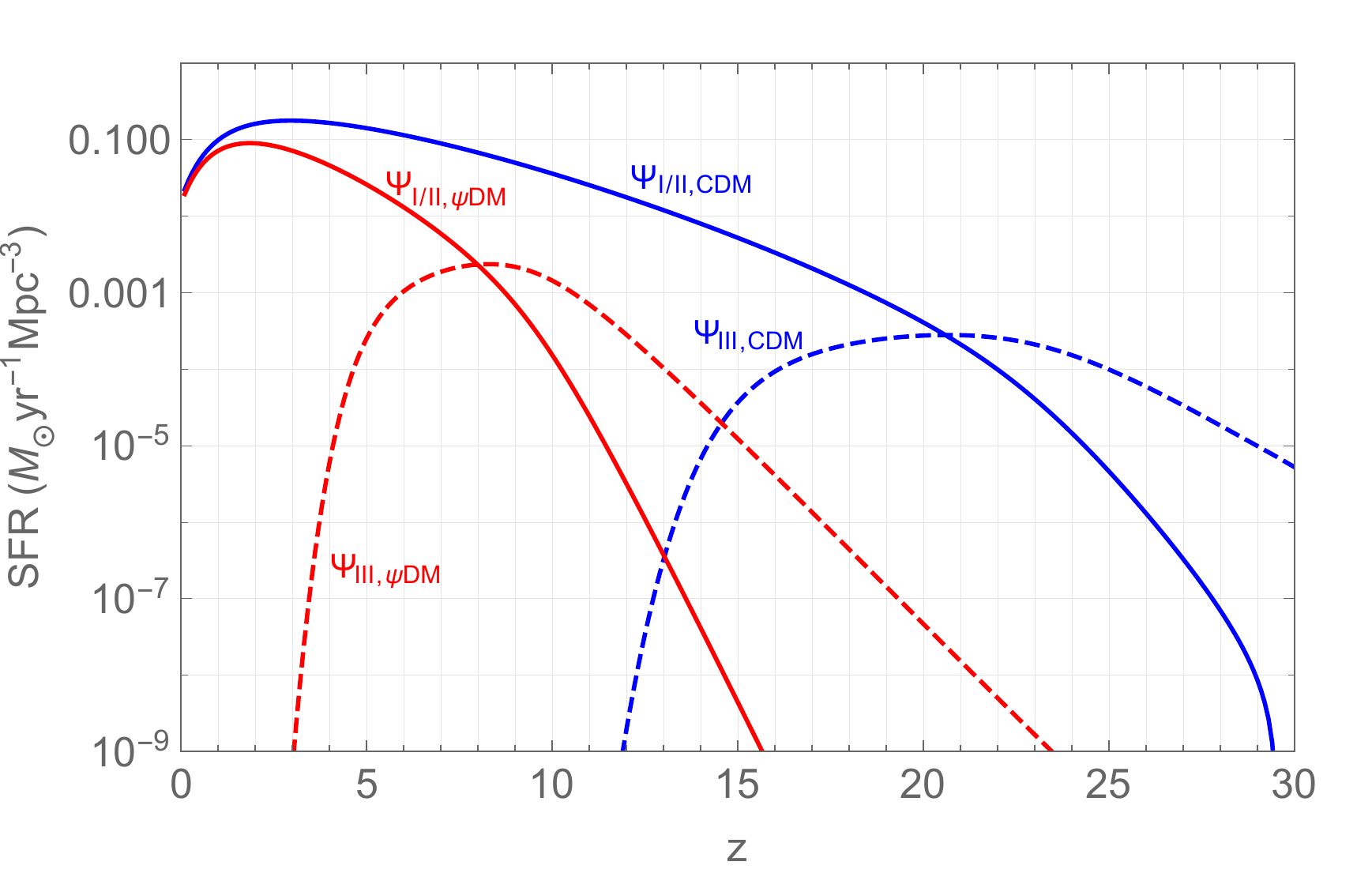}
  \caption{Star formation rate as a function of $z$ for Pop I/II (solid lines) and III stars (dashed lines), assuming CDM (blue lines) and $\psi $DM (red lines) with $m_{22}=1$, respectively. }
  \end{center}
  \label{sfr}
\end{figure}
\section{BH-BH merger rate over the cosmic history}
We restrict our attention to BHs which are produced by the collapse of stars with $25M_\odot\leq m\leq 140M_\odot$ \cite{Tan:2016xvl}, though stars with $260M_\odot\leq m\leq 500M_\odot$ are also expected to collapse to BHs without significant mass loss \cite{Tan:2016xvl}. 
To estimate the merger rate we first calculate BH birth events per unit time per unit comoving volume per unit BH mass by \cite{Dvorkin:2016wac}
\begin{align}
    R_{\mathrm{birth}}(t,m_{\mathrm{bh}})=\int_{25M_\odot}^{100M_\odot}\Psi_{\mathrm{I\hspace{-.1em}/\hspace{-.1em}I\hspace{-.1em}I}}[t-\tau(m)]\phi_{\mathrm{I\hspace{-.1em}/\hspace{-.1em}I\hspace{-.1em}I}}(m)\delta[m-m_r^{-1}(m_{\mathrm{bh}})]dm+\nonumber\\
    \int_{100M_\odot}^{140M_\odot}\Psi_{\mathrm{I\hspace{-.1em}I\hspace{-.1em}I}}[t-\tau(m)]\phi_{\mathrm{I\hspace{-.1em}I\hspace{-.1em}I}}(m)\delta[m-m_r^{-1}(m_{\mathrm{bh}})]dm
\end{align}
for $t>t_i+\tau(25M_\odot)$ and $m_r(25M_\odot)\leq m_{\mathrm{bh}}\leq m_r(140M_\odot)$, where $\delta$ denotes the Dirac delta function. Then, we estimate the BH-BH merger rate by \cite{Dvorkin:2016wac}
\begin{equation}
    R(t)=N\int_{m_r(25M_\odot)}^{m_r(140M_\odot)}dm_{\mathrm{bh}}\int_{t_{\mathrm{d,min}}}^{t_{\mathrm{d,max}}(t)}dt_{d}R_{\mathrm{birth}}(t-t_d,m_{\mathrm{bh}})P(t_d),
\end{equation}
where $P(t_d)\propto t_d^{-1}$, $t_{\mathrm{d,min}}=50$Myr and 
$t_{\mathrm{d,max}}=t-t_i-\tau(25M_\odot)$. $P(t_d)$ is normalized such that $\int_{t_{\mathrm{d,min}}}^{t_{\mathrm{d,max}}(t)}P(t_d)=1$. 
The above expression is valid for $t>t_i+t_{\mathrm{d,min}}+\tau(25M_\odot)$. 
The normalization factor $N$ is chosen \cite{Dvorkin:2016wac} so that $R(z=0)=10^{-7}\mathrm{Mpc}^{-3}\mathrm{yr}^{-1}$ \cite{Abbott:2016nhf}.

BH-BH merger rate as a function of $z$ is shown in Fig.\,2. As expected, the BH merger rate is significantly suppressed at high redshift. More than an order-of-magnitude difference is expected for $z>6$ according to our simplified estimations. This shows that if we observe high-redshift GW events in sufficient numbers, that could provide interesting information about $\psi$DM or other DM and cosmological scenarios where structure formation at high-redshift is modified relative to the standard scenario expected from CDM and nearly-scale-invariant Gaussian primordial fluctuations. According to Fig.\,10 of \cite{Nakamura:2016hna}, the number of events per year per logarithmic redshift interval at $z\sim 6$ can reach $\sim10^5$ for Pre-DECIGO, which implies detections of high-redshift events in future are promising. For sufficiently accurate determinations of the luminosity distance to high-redshift events, we may need to wait for more advanced detectors such as Ultimate-DECIGO. Using such future advanced detectors, we would be able to probe DM physics and early-Universe physics by studying the redshift-distribution of detected GW transient events. If a large stochastic GW background of cosmological and/or astrophysical origins is present in the frequency band of DECIGO, without its successful subtraction it can in principle reduce the signal-to-noise ratio of GWs from merging compact objects, though it is important to note that the DECIGO band is free from the confusion noise caused by numerous white-dwarf binaries \cite{Musha:2017brd}. For DECIGO, if we can characterise the stochastic GW background precisely by cross-correlating data from different units \cite{Musha:2017usi}, thus distinguishing noise and stochastic GWs, then a stochastic GW background wouldn't hinder the detections of high-redshift GW events due to mergers. Cosmic Explorer \cite{Reitze:2019iox} and Einstein Telescope \cite{Maggiore:2019uih}  are also expected to detect BH-BH mergers at high redshift \cite{Chen:2019irf}.
\begin{figure}[htbp]
\begin{center}
  \includegraphics[clip,width=10.0cm]{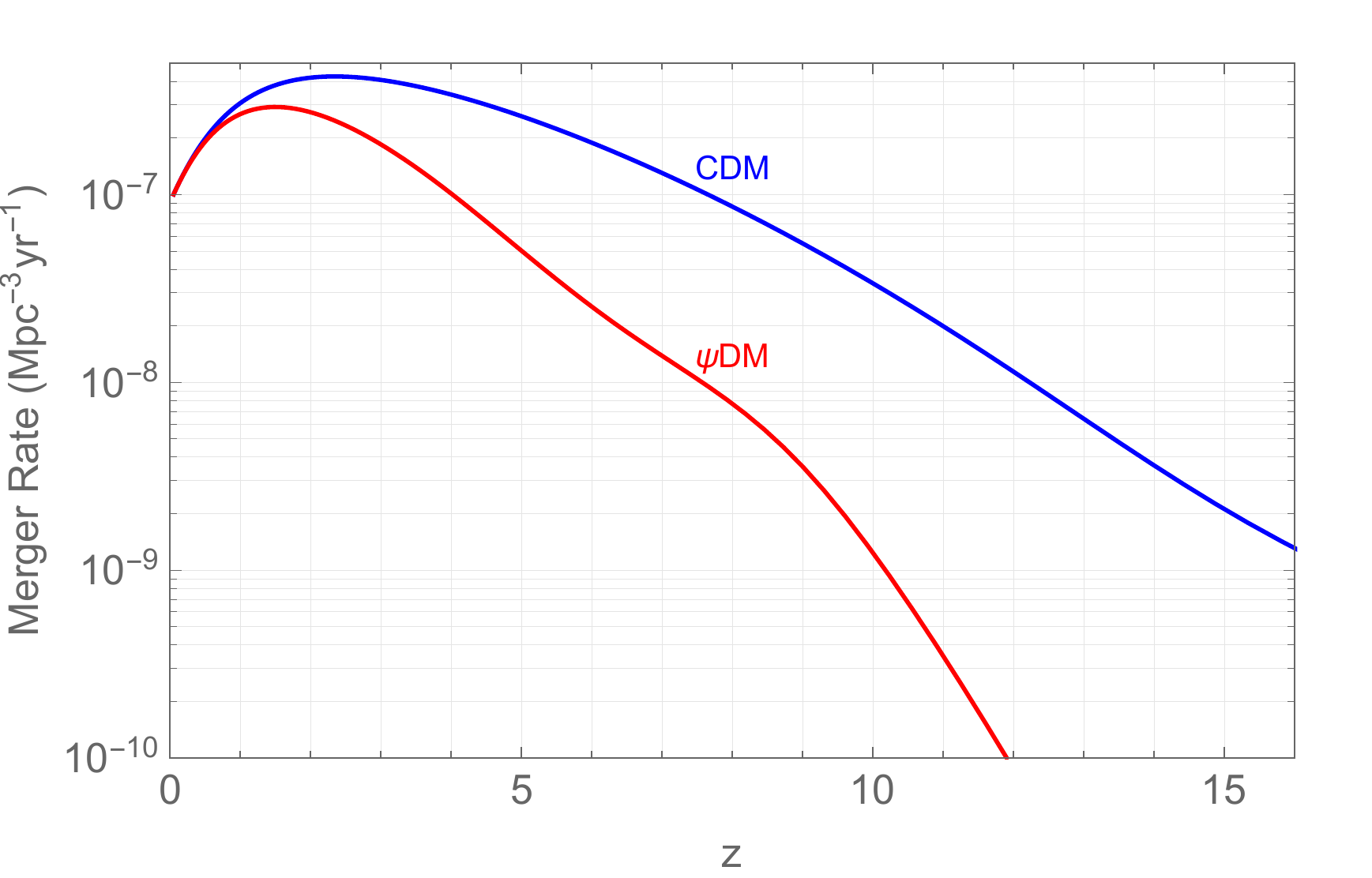}
  \caption{BH-BH merger rate as a function of $z$ for CDM (blue) and $\psi$DM (red) with $m_{22}=1$. }
  \end{center}
\end{figure}

\section{Conclusion and Discussion}
We have shown that the redshift distribution of BH-BH merger rate is sensitive to the structure formation history. Since in future high-redshift GW events are expected to be detected in increasing numbers, we will have enough statistics to understand the redshift distribution of GW events. This would allow us to probe high-redshift structure formation. This implies that we would have additional probes into DM physics and early Universe physics, since high-redshift structure formation is sensitive to them. 

We have used a $\psi$DM scenario as an example, but similar discussions would also apply to other scenarios where structure formation at high redshift significantly differs from the standard CDM predictions under the assumption of the nearly-scale-invariant Gaussian primordial spectrum. They include scenarios where small-scale primordial fluctuations are suppressed due to e.g. inflationary physics, or dark matter physics. These classes of scenarios involving suppression of small-scale power were referred to as \textit{primordial suppression} and \textit{late-time suppression} in \cite{Nakama:2017ohe}. As discusses there, these scenarios have been discussed in the context of $\Lambda$CDM small-scale crisis. Hence, our present work would also have implications for the small-scale crisis. That is, we would be able to know whether some kind of suppression, primordial or late-time, helps resolve the small-scale crisis, or not, in which case the small-scale crisis should be explained solely by astrophysical processes. 

Estimating the merger rate of compact objects involves a number of complex astrophysical processes, and we have adopted simplifying assumptions. Our analysis, we believe, is sufficient for our purpose here, which is to illustrate how the GW event rate as a function of redshift is sensitive to structure formation history, which encodes information about early Universe physics and DM physics. Nonetheless, it would be important and interesting to refine our analysis based on more realistic assumptions, which would make calculations more complicated. Running cosmological simulations, simultaneously resolving astrophysical processes, for different cosmological scenarios would be challenging at this time, so we would need to combine simulations dedicated to each ingredient needed for our estimations and semi-analytic calculations, possibly calibrated by numerical simulations. Recent simulations of \cite{Mocz:2019uyd} report star formation history for different DM scenarios, so our calculations may be improved with the aid of such simulations. Quantifying the relation between SFR and the GW events would also be refined by modeling the details of binary formation as well as the binary evolutions. One may assume GW-driven binary evolution, and also conduct Monte-Carlo simulations to take into account a variety of properties of binaries \cite{Mangiagli:2019sxg}. We may also make use of the methods of \cite{Toffano:2019ekp}, which combined cosmological and astrophysical simulations, to estimate the merger rate more reliably. 

As we have shown, GW experiments would provide additional probes into the early Universe and DM physics. In other words, they will provide additional constraints on early Universe scenarios and DM physics. Such constraints are expected to contain uncertainties stemming from complex astrophysical processes, but in future relevant astrophysical processes are expected to be understood with increasing precision, with the advancement of theoretical, numerical and observational studies. Hence in future such uncertainties would diminish, in which case GW events could provide valuable and reliable tools for us to better understand the early Universe and/or DM physics. 

We have restricted attention to BH-BH mergers, but one could also include other kinds of merger events such as those involving neutron stars. Structure formation history is also in principle encoded in the stochastic GW background, which may also be worth exploring.

\begin{acknowledgments}
The authors would like to thank an anonymous referee for helpful comments.
\end{acknowledgments}


\end{document}